\documentclass[a4paper,fleqn,usenatbib]{mnras}

\usepackage[T1]{fontenc}
\usepackage{ae,aecompl}
\usepackage{graphicx}	
\usepackage{amsmath}	
\usepackage{amssymb}	
\usepackage{cleveref}   
\usepackage{hyperref}   

\title[]{X-ray Intraday Variability of the TeV Blazar Mrk 421 with {\bf \it {Chandra}}}

\author[V. Aggrawal et al.]
{Vishi \ Aggrawal$^{1}$\thanks{E-mail: vishiaggrawal.12@gmail.com}, 
Ashwani Pandey$^{1,2}$\thanks{E-mail: ashwanitapan@gmail.com}, 
Alok C. Gupta$^{1}$\thanks{E-mail: acgupta30@gmail.com},
Zhongli Zhang$^{3}$\thanks{E-mail: zzl@shao.ac.cn}, 
\newauthor
Paul J. Wiita$^{4}$\thanks{E-mail: wiitap@tcnj.edu},
K. K. Yadav$^{5}$,
and S. N. Tiwari$^{2}$
\\ 
\\
$^{1}$Aryabhatta Research Institute of Observational Sciences (ARIES), Manora Peak, Nainital -- 263002, India\\
$^{2}$Department of Physics, DDU Gorakhpur University, Gorakhpur -- 273009, India\\
$^{3}$Shanghai Astronomical Observatory, Chinese Academy of Sciences, 80 Nandan Road, Shanghai 200030, China\\
$^{4}$Department of Physics, The College of New Jersey, 2000 Pennington Road, Ewing, NJ 08628-0718, USA\\
$^{5}$Astrophysical Sciences Division, Bhabha Atomic Research Centre, Mumbai -- 400085, India\\
}

\date{Accepted XXX. Received YYY; in original form ZZZ}

\pubyear{2018}

\begin{document}
\label{firstpage}
\pagerange{\pageref{firstpage}--\pageref{lastpage}}
\maketitle


\begin{abstract}

We present an extensive study of 72 archival {\it Chandra} light curves of the high-frequency-peaked type blazar Mrk 421, the first strong extragalactic object to be detected at TeV energies.
Between 2000 and 2015 Mrk 421 often displayed intraday variability in the 0.3-10.0 keV energy range, as quantified through fractional variability amplitudes that range up to 21.3 per cent.  A variability duty cycle of $\sim$ 84 per cent is present in these data. Variability timescales, with values ranging from 5.5 to 30.5 ks, appear to be present in seven of these 
observations. 
Discrete correlation function analyses show positive correlations between the soft (0.3--2.0 keV) and hard (2.0--10.0 keV) X-ray energy bands with zero time lags, indicating that very similar electron populations are responsible for the emission of all the X-rays observed by {\it Chandra}. The hardness ratios of this X-ray emission indicate a general ``harder-when-brighter" trend in the spectral behaviour of Mrk 421. Spectral index--flux plots provide model independent indications of the spectral evolution of the source and information on the X-ray emission mechanisms. Brief discussions of theoretical models that are consistent with these observations are given.
\end{abstract}



\newpage 
\section{INTRODUCTION} \label{sec:introduction}

An active galactic nucleus (AGN) involves a supermassive black hole (SMBH), fueled by an accretion disc, 
producing a variety of highly energetic phenomena \citep{1984ARA&A..22..471R}. When a radio-loud AGN is 
viewed with one of its relativistic jets in close proximity ($\leq$10$^{\circ}$) to our line of sight, it 
is categorized as a blazar  \citep{1995PASP..107..803U}. Blazars club together BL Lacertae objects, which 
have nearly featureless optical 
continua and many flat spectrum radio quasars, (FSRQs) that show extensive broad emission lines \citep[e.g.,][]{ 2015MNRAS.451.3882A}. 
Blazars are observed to be particularly violent AGNs, involving multiple 
outstanding attributes, including: dominance of non-thermal emission; high polarization;
extreme flux variability across the entire electromagnetic (EM) spectrum; core-dominated radio 
morphology; and flat radio spectrum. All of these can be understood  in terms of relativistic motion of plasma in the jets and
Doppler boosting \citep[e.g.,][]{2016MNRAS.455..680A, 2012AJ....143...23G}. 
The high polarization $( > 3 \%)$ of their radio to 
optical emission means that the synchrotron emission mechanism is responsible for broadband non-thermal EM radiations 
from blazars at lower frequencies (radio through the UV or X-ray bands), while at higher frequencies it is likely to be 
dictated by inverse Compton (IC) scattering of seed photons by the same electrons producing the synchrotron emission. Blazar spectral energy distributions (SEDs) demonstrate a double-peak structure 
\citep[e.g.,][]{2012AJ....143...23G}. The low energy peaked blazars (LBLs) have the first SED bump peak in mm to optical bands 
and the second bump at GeV energies, while the high energy peaked blazars (HBLs) have the first component peak at UV/X-ray 
while the second ranges up to TeV energies \citep{Finke14}. 
The BL Lac/FSRQ sub-classes also can be distinguished based on optical polarization properties: BL Lac objects show 
an amplified 
polarization towards the blue, probably arising due to some intrinsic phenomenon related to the jet-emitting region 
\citep{1991ApJS...76..813S, 1996MNRAS.281..425M}, while the FSRQs trend in the opposite direction, 
possibly because of significant contributions from the unpolarized quasi-thermal emission from the accretion disc 
and surrounding region.

\par Blazar observations often show detectable flux variations down to 
time periods of a few minutes to hours; these must arise from acute physical conditions  within small, subparsec scale, 
regions \citep[e.g.,][]{2016MNRAS.458.1127G}. Blazar variability is conveniently sectioned into three classes, based on their  
observed
time-scales: flux changes occurring over a time-scale of a day or less and up to a few hundredths of a magnitude is 
termed as intra-day 
variability (IDV) \citep{1995ARA&A..33..163W}, or microvariability \citep{1989Natur.337..627M}, or intra-night variability \citep{1993MNRAS.262..963G}; variations in flux, 
typically of
a few tenths of a magnitude, that extend from days to weeks are known as short-term variability (STV); while variations 
ranging from several months 
to a few years are called long-term variability (LTV) \citep{2004A&A...422..505G}. Extensive studies of STV and LTV for 
blazars have often shown variations exceeding $\sim$ 1 mag and some have spanned over $\sim$ 5 mag. These flux 
variabilities in blazars could either be initiated through unstable accretion disc phenomena or solely through changes 
in the doppler-boosted 
emission of the relativistic jets \citep{1997ARA&A..35..445U}. Studies of variability timescales and amplitudes serve as key tools
 in understanding physical processes in the jets and the sizes and locations of the emission regions in AGN.

\subsection{\it Mrk 421}

Markarian 421 (B2 1101+38; Mrk 421 hereafter) is a nearby elliptical active galaxy ($\alpha_{2000}$ = 11h 04m 27.3139s and 
$\delta_{2000}$ = +38$^{\circ}$ 12$\arcmin$ 31.7991$\arcsec$) with an intense point-like 
nucleus, encompassing a $\sim$ 3.6 $\times$ $10^{8}$ M$_{\odot}$ black hole \citep{2008AIPC.1085..399W}. The nuclear
source is classified as of the BL Lacertae type as it has a featureless optical spectrum, 
 strongly polarized and variable optical and radio fluxes, and compact radio emission. Mrk 421 
has a SED well characterised by a classic two peak shape \citep{1995PASP..107..803U, 
1997ARA&A..35..445U}. Most of these observed properties of Mrk 421 are understood to arise from a relativistic jet 
spotted at a small angle to our line of sight \citep{1995PASP..107..803U}.  
Relativistic electrons radiating via the synchrotron process produce a non-thermal SED with a polarised continuum extending 
from the radio to the soft X-ray bands.

\par Mrk 421 ($z=0.031$) is one of the closest blazars, at a distance of 134 Mpc (H$_{0} =71$ km s$^{-1}$ Mpc$^{-1}$, 
$\Omega_{m}$=0.27, $\Omega_{\Lambda}$=0.73) and its synchrotron emission peak was long ago found to lie in the range of 0.1 keV to several 
keV \citep{1992Natur.358..477P}.  The Whipple Cherenkov Telescope claimed to have detected this extragalactic 
source at TeV energy range (0.5--1.5 Tev) \citep{1993ICRC....1..409S} and it has been confirmed as a TeV source by 
multiple ground-based $\gamma$-ray telescopes \citep[e.g.][]{2017ApJ...841..100A}.
The {\it Compton Gamma-Ray Observatory} (CGRO) easily observed Mrk 421 in the GeV band from space.    Mrk 421 is the brightest extragalactic object
in $\gamma-$rays in the northern hemisphere \citep[e.g.,][]{2012AJ....143...23G}.

\par Thanks to its proximity, observational studies of Mrk 421 are pervasive throughout the entire EM spectrum. The source has 
had its radio emission followed
 over the span of 25 years at multiple frequencies \citep[and references therein]{2015MNRAS.448.3121H}. It has shown rapid and extreme optical variability, including 
LTV of $\sim$ 4.6 mag \citep{1976ARA&A..14..173S}, and IDV up to $\sim$ 1.4 mag of brightness change over a very short period ($\sim$ 2.5 hours) \citep{1988A&AS...72..163X}. Three decades of NIR data reported by \cite{1999ApJS..121..131F} provide IDV and STV confirmation of its 
blazar nature.
In 2006, the source was observed with a peak flux $\sim$ 85 mCrab in the 2.0--10.0 keV band, indicating that the first peak 
of SED occurred at an energy beyond 10 keV \citep{2009A&A...501..879T, 2009ApJ...699.1964U}. There were 
reports of ``orphan flares" \citep{2015arXiv150801438F} in TeV $\gamma$-rays, (those not having corresponding increased X-ray emission), 
 in Mrk 421 during 2003 and 2004 multi-wavelength campaigns. On June 10, 2008 Super-AGILE detected a hard X-ray 
flare. 
MAXI (Monitor of ALL-sky X-ray Image) 
marked the strongest X-ray flare in February 2010 ($\sim$ 164$\pm$17 mCrab) \citep{2015ApJ...798...27I}. HESS (High Energy Stereoscopic System) \citep{2005A&A...430..865A} and MAGIC (Major Atmospheric Gamma Imaging Cherenkov Telescopes) \citep{2007ApJ...669..862A} observed the 
time-average high energy spectrum of Mrk 421 during its flaring stages. 

\par The dominant synchrotron-self-Compton (SSC) model considers that the same electron population is responsible for the production 
of soft X-rays and high energy $\gamma$-rays.  The SSC model agrees with the results of \cite{2005ApJ...630..130B} and 
\cite{2009ApJ...695..596H}, where the fluxes were well correlated with a time lag of less than 1.5 days \citep{2016arXiv160509017M}. In 
April 2013, Mrk 421 was observed to undergo a major X-ray outburst and was comprehensively investigated by multiple
observational facilities, including the {\it NuSTAR} and {\it Swift} satellites. Intensive studies of Mrk 421 probing its 
multi-wavelength (MW) behaviour are numerous \citep[and references therein]{1994IAUC.5993....2T, 1995ApJ...438L..59K, 
2004A&A...422..505G, 2008ATel.1574....1C, 2008ATel.1583....1P, 2008ICRC....3..973S, 2012A&A...545A.117L, 2012MNRAS.420.3147G, 
2013A&A...559A..75B, 2014A&A...570A..77P, 2015ApJ...811..143P, 2015A&A...580A.100S, 2015MNRAS.448.3121H, 2016ApJ...819..156B}.
MW campaigns incorporating {\it Fermi-LAT} gamma-ray detections comprehensively studied Mrk 421 and produced
its first ever complete $\gamma$-ray continuum during a quiescent state \citep{2014ApJ...782..110A}. A multi-decade optical light 
curve spanning 1900 to 1991 was extracted by \cite{1997A&AS..123..569L} in the B-band, suggesting two possible observed
time periods of 23.1$\pm$1.1 yrs and 15.3$\pm$0.7 yrs in those flux variations \citep{2016A&A...591A..83S}. A significant 
amount of correlation ($\sim$ 68$\%$) was found with X-ray data from {\it RXTE-ASM}, when \cite{2010tsra.confE.197T} 
studied the long term VHE light curve of the source. Strong episodes of TeV--Xray correlation  were discussed by 
\cite{2005A&A...433..479K}. The typical nature of moderate X-ray--GeV flux correlations has been recently examined
by \cite{2015ApJ...806...20B} through multi-wavelength observations made from 2008 to 2013. 

\par The least well understood aspect of blazar variability is probably seen on IDV timescales.
To search for and analyze IDV in blazars, we are working on a  project in which we study data taken with various
ground and space based telescopes \citep{2008AJ....135.1384G, 2008AJ....136.2359G, 2012MNRAS.425.1357G,
2016MNRAS.458.1127G, 2017MNRAS.465.4423G, 2010ApJ...718..279G, 2012AJ....143...23G, 2012MNRAS.420.3147G, 
2012MNRAS.425.3002G, 2015MNRAS.452.4263G, 2012MNRAS.424.2625B, 2015MNRAS.450..541A, 2015MNRAS.451.3882A, 
2016MNRAS.455..680A, 2015MNRAS.451.1356K, 2017ApJ...841..123P}. In this paper, we present a study of the IDV of Mrk 421
study using the {\it Chandrasekhar X-ray Observatory} satellite. We employ all  the archival data taken by
{\it Chandra} since its launch, extending from 2000 May 29 to 2015 July 02 ($\approx$ 16 years) and totaling 72 IDV light curves.This is the most extensive IDV study of Mrk 421 in the X-ray band, covering the longest temporal span. This work provides us with better understanding of the X-ray variability properties of Mrk 421, along with the correlations 
between hard and soft X-ray bands.

\begin{table*}
Table 1. Observation log of {\it Chandra} data for Mrk 421.
\centering 
\noindent
\scalebox{1.2}{
\begin{tabular}{rcccccccc} 
\hline
\hline
ObsID  &  Date of Observation  &  Start Time (UT)  &  Detector  & Grating &  Exposure Time    \\
       &  (dd-mm-yyyy)         &   (hh:mm:ss)      &            & 	  &   (ks)          \\ \hline
1714 & 29-05-2000  & 11:39:48   & ACIS-S  & HETG   &  19.83    \\ 
1715 & 29-05-2000  & 17:40:11   & HRC-S  & LETG   &  19.84    \\ 
4148 & 26-10-2002  & 00:05:02   & ACIS-S  & LETG   &  96.84    \\ 
4149 & 01-07-2003  & 14:19:35   & HRC-S  & LETG   &  99.98    \\ 
5318 & 06-05-2004  & 14:53:40   & ACIS-S  & LETG   &  30.16    \\ 
5171 & 13-07-2004  & 17:19:41   & ACIS-S  & LETG   &  67.15    \\ 
5332 & 14-07-2004  & 12:21:48   & ACIS-S  & LETG   &  67.06    \\ 
8378 & 07-01-2007  & 23:08:15   & ACIS-S  & LETG   &  28.16    \\ 
6925 & 08-01-2007  & 07:12:48   & ACIS-S  & LETG   &  27.69    \\ 
8396 & 21-01-2007  & 07:52:59   & HRC-S  & LETG   &  29.48    \\ 
10671 & 10-10-2009  & 23:53:51   & ACIS-S  & LETG   &  28.96    \\ 
10664 & 08-11-2009  & 05:00:56   & ACIS-S  & LETG   &  20.06    \\ 
11605 & 16-11-2009  & 03:43:14   & ACIS-S  & LETG   &  ~5.24    \\ 
11606 & 18-11-2009  & 19:16:47   & ACIS-S  & LETG   &  ~5.24    \\ 
11607 & 22-11-2009  & 09:17:23   & ACIS-S  & LETG   &  ~5.14    \\ 
11960 & 02-02-2010  & 03:31:10   & ACIS-I  & LETG   &  20.13    \\ 
11961 & 04-02-2010  & 08:50:00   & ACIS-I  & LETG   &  20.17    \\ 
11962 & 04-02-2010  & 14:44:43   & ACIS-I  & LETG   &  19.77    \\ 
11963 & 06-02-2010  & 08:54:32   & ACIS-I  & LETG   &  20.18    \\ 
11964 & 06-02-2010  & 14:56:02   & ACIS-I  & LETG   &  19.77    \\ 
11967 & 06-02-2010  & 20:37:52   & ACIS-I  & LETG   &  19.78    \\ 
10663 & 13-03-2010  & 02:11:06   & ACIS-S  & HETG   &  15.13    \\ 
11970 & 13-03-2010  & 06:56:51   & ACIS-S  & LETG   &  10.07    \\ 
10665 & 13-03-2010  & 10:08:56   & HRC-S  & LETG   &  10.15    \\ 
12121 & 13-03-2010  & 13:07:43   & ACIS-S  & LETG   &  10.07    \\ 
10667 & 13-03-2010  & 16:14:18   & HRC-S  & LETG   &  10.18    \\ 
10668 & 13-03-2010  & 19:13:05   & ACIS-S  & LETG   &  10.06    \\ 
10669 & 13-03-2010  & 22:19:39   & HRC-S  & LETG   &  10.18    \\ 
11966 & 14-03-2010  & 01:18:27   & ACIS-S  & LETG   &  30.06    \\ 
10670 & 14-03-2010  & 10:00:29   & ACIS-S  & HETG   &  14.80    \\ 
12122 & 10-07-2010  & 18:37:09   & HRC-S  & LETG   &  25.18    \\ 
13097 & 16-02-2011  & 18:33:05   & ACIS-S  & LETG   &  30.06    \\ 
13098 & 04-07-2011  & 02:15:37   & ACIS-S  & HETG   &  14.80    \\ 
13099 & 04-07-2011  & 06:59:47   & ACIS-S  & LETG   &  10.06    \\ 
13100 & 04-07-2011  & 10:12:04   & HRC-S  & LETG   &  10.15    \\ \hline          
\end{tabular}}
\end{table*}

\begin{table*}
Table 1. continued.

\noindent
\scalebox{1.2}{
\begin{tabular}{lcccccccc} \hline
\hline
ObsID  &  Date of Observation  &  Start Time (UT)  &  Detector  & Grating &  Exposure Time    \\
       &  (dd-mm-yyyy)         &   (hh:mm:ss)      &            & 	  &   (ks)          \\ \hline
13104 & 04-07-2011  & 22:07:18   & HRC-S  & LETG   &  10.18    \\ 
13105 & 05-07-2011  & 01:06:06   & ACIS-S  & HETG   &  15.00    \\ 
14266 & 07-04-2012  & 20:29:18   & ACIS-S  & LETG   &  30.05    \\ 
14320 & 03-07-2012  & 11:34:05   & ACIS-S  & HETG   &  15.03    \\ 
14322 & 03-07-2012  & 16:07:49   & HRC-S  & LETG   &  10.01    \\ 
14396 & 03-07-2012  & 19:09:24   & HRC-S  & LETG   &  ~9.79    \\ 
14321 & 03-07-2012  & 22:06:42   & ACIS-S  & LETG   &  10.05    \\ 
14323 & 04-07-2012  & 01:13:05   & HRC-S  & LETG   &  10.18    \\ 
14324 & 04-07-2012  & 04:09:45   & HRC-S  & LETG   &  ~9.79    \\ 
14325 & 04-07-2012  & 07:07:03   & ACIS-S  & LETG   &  10.05    \\ 
14326 & 04-07-2012  & 10:13:26   & HRC-S  & LETG   &  10.19    \\ 
14397 & 04-07-2012  & 13:10:06   & HRC-S  & LETG   &  ~9.79    \\ 
14327 & 04-07-2012  & 16:07:26   & ACIS-S  & HETG   &  15.04    \\ 
15607 & 07-02-2013  & 19:15:19   & ACIS-S  & LETG   &  30.07    \\ 
15476 & 03-04-2013  & 00:36:24   & ACIS-S  & LETG   &  30.05    \\ 
15477 & 30-06-2013  & 16:15:37   & ACIS-S  & HETG   &  14.65    \\ 
15478 & 30-06-2013  & 20:46:21   & ACIS-S  & LETG   &  10.06    \\ 
15479 & 30-06-2013  & 23:58:37   & HRC-S  & LETG   &  10.14    \\ 
15480 & 01-07-2013  & 02:57:24   & ACIS-S  & LETG   &  ~9.76    \\ 
15481 & 01-07-2013  & 05:56:13   & HRC-S  & LETG   &  10.19    \\ 
15482 & 01-07-2013  & 08:55:01   & ACIS-S  & LETG   &  ~9.77    \\ 
15483 & 01-07-2013  & 11:53:49   & HRC-S  & LETG   &  10.19    \\ 
15484 & 01-07-2013  & 14:52:38   & ACIS-S  & HETG   &  14.50    \\ 
16474 & 06-03-2014  & 08:13:12   & ACIS-S  & LETG   &  60.07    \\ 
16424 & 25-06-2014  & 13:54:38   & ACIS-S  & HETG   &  15.05    \\ 
16425 & 25-06-2014  & 20:44:18   & ACIS-S  & LETG   &  10.08    \\ 
16426 & 25-06-2014  & 23:56:09   & HRC-S  & LETG   &  10.15    \\ 
16427 & 26-06-2014  & 02:55:01   & ACIS-S  & LETG   &  10.08    \\ 
16428 & 26-06-2014  & 06:01:20   & HRC-S  & LETG   &  10.18    \\ 
16429 & 26-06-2014  & 09:00:09   & ACIS-S  & LETG   &  10.08    \\ 
16430 & 26-06-2014  & 12:06:31   & HRC-S  & LETG   &  10.17    \\ 
16431 & 26-06-2014  & 15:05:17   & ACIS-S  & HETG   &  15.05    \\ 
17385 & 01-07-2015  & 00:42:11   & ACIS-S  & HETG   &  15.03    \\ 
17387 & 01-07-2015  & 20:54:30   & HRC-S  & LETG   &  10.18    \\ 
17389 & 01-07-2015  & 23:51:09   & HRC-S  & LETG   &  10.19    \\ 
17391 & 02-07-2015  & 02:46:19   & HRC-S  & LETG   &  10.18    \\ 
17392 & 02-07-2015  & 05:43:39   & ACIS-S  & HETG   &  14.06    \\  \hline
\end{tabular}}
\end{table*}

\par The paper is organized as follows. Section 2 briefly describes the {\it Chandra} satellite instrumentation along with the methodology 
for data reduction. Data analysis techniques used to search for flux and spectral variability are discussed in 
Section 3. Section 4 and Section 5 give our results and a discussion, respectively. Our conclusions are reported in 
Section 6.

\section{{\it Chandra} X-RAY SATELLITE AND DATA REDUCTION} \label{sec:chandra}

Launched on July 23, 1999 as a part of the ``Great Observatories", this {\it NASA} telescope has been a phenomenal tool to study high energy sources such as compact binaries, quasars and supernovas. {\it Chandra's} High Resolution Mirror Assembly (HRMA) typically produces images with a half-power diameter (HPD) of the point spread function (PSF) of $<$ 0.5 arcsec. The High Energy Transmission Grating (HETG) and the Low Energy Transmission Grating (LETG) have high resolving powers when compared to their bandwidths and together cover the energy range from $\leq$0.1 to 10 keV. ACIS (Advanced CCD Imaging Spectrometer) and HRC (High Resolution Camera) are the two in-house focal instruments of the Science Instrument Module (SIM) \citep[e.g.,][]{2000SPIE.4012....2W}.

\par ACIS consists of two CCD arrays: four arrangements of 2$\times$2 arrays, known as ACIS-I (having front-illuminated (FI) CCDs) and six arrangements of 1$\times$6 arrays, ACIS-S (consisting of 4 FI and 2 back-illuminated (BI) CCDs). The time resolution of two ACIS detectors is 3.2 sec \citep[e.g.,][]{2000SPIE.4012....2W}. When observing a wide-field (16$\arcmin$$\times$16$\arcmin$) and/or requiring high energy response, ACIS-I is preferred while imaging observations having a low energy response and a smaller field-of-view (8$\arcmin$$\times$8$\arcmin$) are provided by ACIS-S. 

\par The micro channel plate instrument, HRC, has the fastest time resolution of 16 $\mu$sec.  It employs two detectors, one of which, HRC-I is calibrated for imaging with a wide field-of-view (FoV) of $\sim 30^{\prime} \times 30^{\prime}$.  The other, HRC-S, is primarily used with the LETG and has a long FoV of $\sim 7^{\prime} \times 97^{\prime}$ \citep[e.g.,][]{2000SPIE.4012....2W}.

\subsection{\it Data Reduction}

Mrk 421 was observed by {\it Chandra} between 2000 May 29 to  2015 July 02, providing a rich set of data for the study of its variable nature in the X-ray energy range of 0.3-10 keV.   We downloaded  72 observation IDs from the HEASARC Data Archive\footnote{\url{https://heasarc.gsfc.nasa.gov/db-perl/W3Browse/w3browse.pl}}. The list of  Observation IDs, dates, start times, detectors used, gratings and exposure times are given in Table 1.
The majority of these observations were used for calibration or spectroscopic study of the warm-hot intergalactic medium filaments in the direction of the blazar 
\citep[e.g.,][]{2005ApJ...629..700N, 2006ApJ...652..189K, 2007ApJ...656..129R, 2008ApJ...672L..21Y}.  Hence the extensive timing analysis we present here is unique for {\it Chandra} data.

\par The {\it Chandra} Interactive Analysis of Observations (CIAO\footnote{\url{http://cxc.harvard.edu/ciao/}} version 4.9) package was used in conjunction with CALDB version  4.7.7 to process the data. 
We first reprocessed the level 2 event file to apply the updated calibration data using the CIAO script {\it chandra\_repro}. We then applied a barycenter correction to the reprocessed level 2 event file using the CIAO tool {\it axbary}. Out of 72 observations, 48 were done with the ACIS detector and remaining 24 with HRC. Although gratings are used in all these observations, there is a possibility of pile-up in the undispersed (zeroth order) events for a bright source such as Mrk 421 in observations performed with the ACIS detector. The dispersed (first order) events are, however, free from pile-up. So, for the 48 observations taken with ACIS, we determined fluxes from a rectangular region of  $800^{\prime\prime} \times 20^{\prime\prime}$ that contains only the dispersed source photons. For observations made with HRC we took a circular region of radius $10^{\prime\prime}$ centered on the source to extract light curves.
We have also taken into account the Dead Time Factor (DTF\footnote{The DTF characterises the detector's deviation from the standard detection efficiency.}) while creating the HRC light curves. 
Finally the 0.3--10.0 keV light curves were extracted using the CIAO tool {\it dmextract} with a binning of 500 secs. Due to the brightness of Mrk 421, the background contribution is negligible. 
 
\begin{table*}
Table 2. X-ray variability parameters.
\centering
\noindent
\scalebox{1.2}{
\begin{tabular}{lcccccc} \hline
\hline
 
ObsID  &                      \multicolumn{3}{c} {F$_{var}$(per cent)}        &  ACF     &   Bin-size \\
\cline{2-4}       

       &                  Soft(0.3-2 keV)	 &Hard(2-10 keV)	    &Total(0.3-10 keV)       &  (ks)    &     (ks)  \\ \hline
~1714     & $   15.24 \pm   0.54  $ &  	    $   18.96 \pm   0.54 $   	  &  $   16.79 \pm   0.37 $  &  $      	-        $& 1.00 \\ 
~1715$^*$     & $  	        -        $ &  	    $  	        -  	 $   	  &  $   ~2.62 \pm   0.30 $  &  $      	-        $& 1.00 \\ 
~4148     & $   10.49 \pm   0.06  $ &  	    $   16.27 \pm   0.09 $   	  &  $   12.27 \pm   0.05 $  &  $      30.50     $& 3.00 \\ 
~4149$^*$     & $  	        -        $ &  	    $  	        -  	 $   	  &  $   18.56 \pm   0.07 $  &  $      	-        $& 5.00 \\ 
~5318     & $   ~9.51 \pm   0.12  $ &  	    $   14.62 \pm   0.16 $   	  &  $   11.02 \pm   0.09 $  &  $      19.50     $& 1.50 \\ 
~5171     & $   ~6.02 \pm   0.12  $ &  	    $   ~9.73 \pm   0.19 $   	  &  $   ~7.03 \pm   0.10 $  &  $      	-        $& 5.00 \\ 
~5332     & $   ~4.35 \pm   0.12  $ &  	    $   ~8.67 \pm   0.21 $   	  &  $   ~5.12 \pm   0.10 $  &  $      	-        $& 3.00 \\ 
~8378     & $   ~3.77 \pm   0.24  $ &  	    $   ~4.46 \pm   0.49 $   	  &  $   ~3.95 \pm   0.21 $  &  $      	-        $& 2.00 \\ 
~6925     & $   ~2.15 \pm   0.27  $ &  	    $   ~4.43 \pm   0.49 $   	  &  $   ~2.80 \pm   0.24 $  &  $      10.07     $& 1.00 \\ 
~8396$^*$    & $  	        -        $ &  	    $  	        -  	 $   	  &  $   ~7.22 \pm   0.16 $  &  $      	-        $& 2.00 \\ 
10671    & $   ~3.94 \pm   0.17  $ &  	    $   ~5.14 \pm   0.26 $   	  &  $   ~4.32 \pm   0.14 $  &  $      14.98	 $& 1.00 \\ 
10664    & $   ~2.54 \pm   0.18  $ &  	    $   ~2.39 \pm   0.31 $   	  &  $   ~2.21 \pm   0.16 $  &  $      	-        $& 1.00 \\ 
11605    & $   ~1.67 \pm   0.35  $ &  	    $   ~1.66 \pm   0.56 $   	  &  $   ~2.07 \pm   0.32 $  &  $      	-        $& 0.60 \\ 
11606    & $   ~1.11 \pm   0.40  $ &  	    $   ~0.23 \pm   2.66 $   	  &  $   ~0.86 \pm   0.47 $  &  $      	-        $& 0.60 \\ 
11607    & $   ~1.04 \pm   0.45  $ &  	    $   ~3.14 \pm   0.62 $   	  &  $   ~1.41 \pm   0.33 $  &  $      	-        $& 0.70 \\ 
11960    & $   ~7.27 \pm   0.36  $ &  	    $   10.56 \pm   0.53 $   	  &  $   ~8.21 \pm   0.29 $  &  $      	-        $& 2.00 \\ 
11961    & $   ~7.45 \pm   0.26  $ &  	    $   10.55 \pm   0.31 $   	  &  $   ~8.26 \pm   0.19 $  &  $      	-        $& 2.00 \\ 
11962    & $   ~4.16 \pm   0.28  $ &  	    $   ~5.77 \pm   0.39 $   	  &  $   ~4.66 \pm   0.22 $  &  $      	10.07    $& 1.00 \\ 
11963    & $   ~4.08 \pm   0.24  $ &  	    $   ~4.03 \pm   0.30 $   	  &  $   ~3.65 \pm   0.20 $  &  $      	-        $& 2.00 \\ 
11964    & $   ~1.32 \pm   0.37  $ &  	    $   ~3.97 \pm   0.34 $   	  &  $   ~3.13 \pm   0.22 $  &  $      	-        $& 2.00 \\ 
11967    & $   ~5.80 \pm   0.38  $ &  	    $   ~8.11 \pm   0.52 $   	  &  $   ~8.55 \pm   0.30 $  &  $      	-        $& 2.00 \\ 
10663    & $   ~2.33 \pm   0.36  $ &  	    $   ~4.36 \pm   0.30 $   	  &  $   ~3.60 \pm   0.23 $  &  $      	-        $& 0.60 \\ 
11970    & $   ~2.11 \pm   0.28  $ &  	    $   ~4.52 \pm   0.43 $   	  &  $   ~2.48 \pm   0.20 $  &  $      	-        $& 1.00 \\ 
10665$^*$    & $            -        $ &  	    $  	        -  	 $   	  &  $   ~0.59 \pm   0.40 $  &  $      	-        $& 0.60 \\ 
12121    & $   ~7.13 \pm   0.23  $ &  	    $   13.42 \pm   0.37 $   	  &  $   ~8.43 \pm   0.21 $  &  $      	-        $& 1.00 \\ 
10667$^*$ & $  	        -        $ &  	    $  	        -  	 $   	  &  $   ~5.71 \pm   0.22 $  &  $      	-        $& 1.00 \\ 
10668    & $   ~7.84 \pm   0.20  $ &  	    $   10.72 \pm   0.29 $   	  &  $   ~8.45 \pm   0.17 $  &  $      	-        $& 1.00 \\ 
10669$^*$ & $  	        -        $ &  	    $  	        -  	 $   	  &  $   ~2.59 \pm   0.25 $  &  $      	-        $& 1.00 \\ 
11966    & $   18.70 \pm   0.15  $ &  	    $   27.95 \pm   0.22 $   	  &  $   21.31 \pm   0.12 $  &  $      	-        $& 3.00 \\ 
10670    & $   ~2.52 \pm   0.55  $ &  	    $   ~4.37 \pm   0.50 $   	  &  $   ~3.66 \pm   0.35 $  &  $      	-        $& 1.00 \\ 
12122$^*$  & $  	        -        $ &  	    $  	        -  	 $   	  &  $   ~2.54 \pm   0.54 $  &  $      	-        $& 2.00 \\ 
13097    & $   ~6.42 \pm   1.09  $ &  	    $   ~7.93 \pm   2.47 $   	  &  $   ~6.28 \pm   0.95 $  &  $      	-        $& 2.00 \\ 
13098    & $   ~5.13 \pm   0.75  $ &  	    $   ~4.14 \pm   0.81 $   	  &  $   ~4.78 \pm   0.56 $  &  $      	-        $& 1.00 \\ 
13099    & $   ~2.48 \pm   0.51  $ &  	    $   ~2.72 \pm   1.00 $   	  &  $   ~1.65 \pm   0.45 $  &  $      	-        $& 1.00 \\ 
13100$^*$  & $  	        -        $ &  	    $  	        -  	 $   	  &  $   ~2.07 \pm   0.49 $  &  $      	-        $& 1.00 \\ 
\hline
\end{tabular}}
\\$^*$ Observation done with HRC detector in which the light curve can not be split into different energy bands, hence the values of $F_{var}$ are not quoted for Soft and Hard energy bands. 
\end{table*}

\begin{table*}
 Table 2. continued.
\centering

\scalebox{1.2}{
\begin{tabular}{lccccccc} \hline
\hline
ObsID  &                     \multicolumn{3}{c} {F$_{var}$({\it percent})}	     &   ACF    &   Bin-Size \\
\cline{2-4} 

       &                  Soft(0.3-2 keV)	 &Hard(2-10 keV)	    &Total(0.3-10 keV)       &  (ks)    &     (ks)  \\ \hline
13104$^*$    & $  	        -        $ &  	    $  	        -  	 $   	  &  $   ~1.69 \pm   0.48 $  &  $      	-        $& 1.00 \\ 
13105    & $   ~4.21 \pm   0.68  $ &  	    $   ~6.21 \pm   0.65 $   	  &  $   ~4.56 \pm   0.46 $  &  $      	-        $& 2.00 \\ 
14266    & $   ~3.28 \pm   0.41  $ &  	    $   ~4.93 \pm   0.73 $   	  &  $   ~3.33 \pm   0.35 $  &  $    ~5.51       $& 1.00 \\ 
14320    & $   ~3.57 \pm   0.95  $ &  	    $   ~6.11 \pm   0.99 $   	  &  $   ~4.33 \pm   0.66 $  &  $      	-        $& 1.00 \\ 
14322$^*$    & $  	        -        $ &  	    $  	        -  	 $   	  &  $   ~2.68 \pm   0.43 $  &  $      	-        $& 1.00 \\ 
14396$^*$    & $  	        -        $ &  	    $  	        -  	 $   	  &  $   ~0.86 \pm   0.64 $  &  $      	-        $& 1.00 \\ 
14321    & $   ~1.45 \pm   0.59  $ &  	    $   ~2.33 \pm   1.25 $   	  &  $   ~1.06 \pm   0.62 $  &  $      	-        $& 1.00 \\ 
14323$^*$    & $  	        -        $ &  	    $  	        -  	 $   	  &  $   ~2.52 \pm   0.42 $  &  $      	-        $& 1.00 \\ 
14324$^*$    & $  	        -        $ &  	    $  	        -  	 $   	  &  $   ~2.53 \pm   0.40 $  &  $      	-        $& 1.00 \\ 
14325    & $   ~1.21 \pm   0.62  $ &  	    $   ~1.04 \pm   2.11 $   	  &  $   ~0.84 \pm   0.69 $  &  $      	-        $& 1.00 \\ 
14326$^*$    & $  	        -        $ &  	    $  	        -  	 $   	  &  $   ~8.98 \pm   0.34 $  &  $      	-        $& 1.00 \\ 
14397$^*$    & $  	        -        $ &  	    $  	        -  	 $   	  &  $   ~1.19 \pm   0.44 $  &  $      	-        $& 1.00 \\ 
14327    & $   ~5.11 \pm   0.60  $ &  	    $   ~7.86 \pm   0.62 $   	  &  $   ~6.16 \pm   0.44 $  &  $      	-        $& 1.00 \\ 
15607    & $   ~6.87 \pm   0.19  $ &  	    $   10.98 \pm   0.31 $   	  &  $   ~7.65 \pm   0.15 $  &  $    ~9.51       $& 1.00 \\ 
15476    & $   10.05 \pm   0.16  $ &  	    $   16.73 \pm   0.23 $   	  &  $   12.03 \pm   0.13 $  &  $      	-        $& 1.00 \\ 
15477    & $   ~5.44 \pm   0.42  $ &  	    $   ~8.60 \pm   0.40 $   	  &  $   ~7.09 \pm   0.28 $  &  $      	-        $& 1.00 \\ 
15478    & $   ~7.63 \pm   0.30  $ &  	    $   10.60 \pm   0.49 $   	  &  $   ~8.45 \pm   0.22 $  &  $      	-        $& 1.00 \\ 
15479$^*$    & $  	        -        $ &  	    $  	        -  	 $   	  &  $   ~4.52 \pm   0.26 $  &  $      	-        $& 1.00 \\ 
15480    & $   ~1.87 \pm   0.33  $ &  	    $   ~2.86 \pm   0.61 $   	  &  $   ~2.00 \pm   0.29 $  &  $      	-        $& 1.00 \\ 
15481$^*$    & $  	        -        $ &  	    $  	        -  	 $   	  &  $   ~2.52 \pm   0.29 $  &  $      	-        $& 1.00 \\ 
15482    & $   ~0.51 \pm   0.61  $ &  	    $   ~1.40 \pm   0.79 $   	  &  $   ~0.74 \pm   0.43 $  &  $      	-        $& 1.00 \\ 
15483$^*$    & $  	        -        $ &  	    $  	        -  	 $   	  &  $   ~5.07 \pm   0.28 $  &  $      	-        $& 1.00 \\ 
15484    & $   ~3.41 \pm   0.49  $ &  	    $   ~5.44 \pm   0.46 $   	  &  $   ~4.58 \pm   0.31 $  &  $      	-        $& 1.00 \\ 
16474    & $   ~9.59 \pm   0.09  $ &  	    $   12.12 \pm   0.13 $   	  &  $   10.38 \pm   0.07 $  &  $      	-        $& 5.00 \\ 
16424    & $   ~7.77 \pm   0.70  $ &  	    $   12.88 \pm   0.69 $   	  &  $   10.67 \pm   0.48 $  &  $      	-        $& 1.00 \\ 
16425    & $   ~3.29 \pm   0.50  $ &  	    $   ~3.70 \pm   0.91 $   	  &  $   ~3.10 \pm   0.41 $  &  $      	-        $& 1.00 \\ 
16426$^*$    & $  	        -        $ &  	    $  	        -  	 $   	  &  $   ~3.82 \pm   0.40 $  &  $      	-        $& 1.00 \\ 
16427    & $   ~2.09 \pm   0.47  $ &  	    $   ~1.19 \pm   1.30 $   	  &  $   ~2.00 \pm   0.43 $  &  $      	-        $& 1.00 \\ 
16428$^*$    & $  	        -        $ &  	    $  	        -  	 $   	  &  $   ~3.39 \pm   0.41 $  &  $      	-        $& 1.00 \\ 
16429    & $   ~6.65 \pm   0.44  $ &  	    $   ~9.31 \pm   0.74 $   	  &  $   ~6.79 \pm   0.39 $  &  $      	-        $& 1.00 \\ 
16430$^*$    & $  	        -        $ &  	    $  	        -  	 $   	  &  $   ~4.76 \pm   0.41 $  &  $      	-        $& 1.00 \\ 
16431    & $   ~1.59 \pm   1.11  $ &  	    $   ~1.67 \pm   1.06 $   	  &  $   ~2.33 \pm   0.58 $  &  $      	-        $& 1.00 \\ 
17385    & $   ~7.03 \pm   0.75  $ &  	    $   ~7.25 \pm   0.73 $   	  &  $   ~7.43 \pm   0.52 $  &  $      	-        $& 1.00 \\ 
17387$^*$    & $  	        -        $ &  	    $  	        -  	 $   	  &  $   ~3.52 \pm   0.34 $  &  $      	-        $& 1.00 \\ 
17389$^*$    & $  	        -        $ &  	    $  	        -  	 $   	  &  $   ~6.18 \pm   0.29 $  &  $      	-        $& 1.00 \\ 
17391$^*$    & $  	        -        $ &  	    $  	        -  	 $   	  &  $   ~2.43 \pm   0.30 $  &  $      	-        $& 1.00 \\ 
17392    & $   ~4.78 \pm   0.55  $ &  	    $   ~7.57 \pm   0.52 $   	  &  $   ~6.15 \pm   0.37 $  &  $      	-        $& 1.00 \\ \hline
\end{tabular}}
\\$^*$ Observation done with HRC detector in which the light curve cannot be split into different energy bands, hence the values of $F_{var}$ are not quoted for Soft and Hard energy bands. 
\end{table*}

\noindent
\section{ANALYSIS TECHNIQUES} \label{sec:analysis}

\subsection{\it Excess Variance} \label{subsec:frac}
Although AGNs in general, and blazars in particular, are generally characterized by rapid X-ray variability, when observed there will 
be some innate experimental noise. Measurement errors in the LC produce finite uncertainties, $\sigma_{err,i}$, for each of the $i$ measurements that contribute additional variance to the observed variance
\citep[e.g.,][]{2017ApJ...841..123P}. The quantitative measure of the true variance is known as excess variance and yields the magnitude of variability.  
Thus the excess variance is defined for 
an observed LC having $N$ measured flux values $x_i$ as \citep{2003MNRAS.345.1271V}
\begin{equation}
\sigma_{XS}^2 = S^2 - \overline{\sigma_{err}^2},
\end{equation}
where $S^2$ is the total variance of the LC, and is given by
\begin{equation}
S^2 = \frac{1}{N-1} \sum\limits_{i=1}^N (x_i - \bar{x})^2 ,
\end{equation}
and $\overline{\sigma_{err}^2}$ is the mean square error is given by
\begin{equation} 
\overline{\sigma_{err}^2} =\frac{1}{N} \sum\limits_{i=1}^N \sigma^2_{err,i},
\end{equation}
where $\bar{x}$ is the arithmetic mean of $x_i$. \\

It makes sense to normalise the excess variance 
\begin{equation}
\sigma^2_{NXS} = {\sigma^2_{XS}} / {\bar{x}^2},
\end{equation}
and the fractional rms variability amplitude, $F_{var}$, is defined as
\begin{equation}
F_{var} = \sqrt{\frac{S^2 - \overline{\sigma_{err}^2}}{{\bar{x}^2}}},
\end{equation}
It can be shown that the uncertainty in $F_{var}$ is given by \citep[e.g.,][]{2003MNRAS.345.1271V}, 
\begin{equation}
err(F_{var}) =  \sqrt{\left( \sqrt{\frac{1}{2N}}\frac{\overline{\sigma_{err}^2}}{\bar{x}^2 F_{var}} \right)  ^ 2+ \left(  \sqrt{\frac{\overline{\sigma_{err}^2}}{N}} \frac{1}{\bar{x}}\right) ^2 }.
\end{equation}

\subsection{ \it Discrete Correlation Functions}\label{sec:DCF}

The Discrete Correlation Function (DCF) technique was first introduced to astronomical time series by \cite{1988ApJ...333..646E} and later modified to give better 
error estimates by \cite{1992ApJ...386..473H}.  The DCF can be used to find possible time-lags between bands and so correlate multi-frequency AGN 
LCs data, even when they are distributed irregularly over time, as is typical. Two discrete data trains $x$ and $y$ are taken into account to collect 
unbinned correlations (UDCF) \citep[e.g.,][]{2017ApJ...841..123P} via
\begin{equation}
 F_{ij}=\frac{(x_i - \bar{x})(y_j - \bar{y})}{\sqrt{\left(\sigma_x^2\right)\left(\sigma_y^2\right)}},
\end{equation}
for the data sets $x_i$ and $y_j$ having $\bar{x}$ and $\bar{y}$ as their mean values, and $\sigma_x$  and  $\sigma_y$  as their 
standard deviations, respectively \citep{2015A&A...582A.103G}.  

\par In order to obtain a normalized DCF, one must bin the calculated UDCF, but the selected bin size should neither be too large to lose any important data 
points, nor too small so as to produce specious correlations. The DCF then can be computed by taking the average of a number, $M$,  of UDCF 
values for each time delay $\Delta t_{ij} = t_j - t_i$ that lie in the range 
$\tau - \Delta\tau/2  \leq  \Delta t_{ij} < \tau + \Delta \tau/2$ as,
\begin{equation}
DCF(\tau)=\frac{1}{M}\sum  F_{ij},
\end{equation}
with $\tau$ being the center of the time bin $\Delta\tau$. Each bin has a standard error estimate that is given as 
\citep{1988ApJ...333..646E},
\begin{equation}
\sigma_{DCF}(\tau) = \frac{\sqrt{\sum[F_{ij} - DCF(\tau)]^2}}{M - 1},
\end{equation}

The Auto Correlation Function (ACF) is the special case of the DCF where a data train is correlated with itself ($x = y$), automatically
generating a peak at $\tau = 0$. This peak points out the absence of any time lag, and any periodicity in the data would be marked 
by the presence of other strong peaks \citep{2015MNRAS.450..541A}. Generally, two correlated data signals have a DCF 
value $>$ 0, while a DCF value $<$ 0 indicates that the two data sets are anti-correlated, and a DCF value $=$ 0 means that 
the two data trains have no  correlation \citep{2014ApJ...781L...4G}.

\subsection{ \it Hardness Ratio}\label{sec:HR}

The Hardness Ratio (HR) is an effective and elementary model-independent method which can simply characterise spectral variations, and is defined as
\begin{equation}
HR = \frac{(H-S)}{(H+S)},
\end{equation}
where, $H$ is the net count rate in the hard energy band and $S$ is the net count rate in the soft energy band 
\citep[e.g.,][]{2016ApJ...824...51P}. In order to examine spectral variability in our X-ray light curves, we split the LCs into a soft X-ray energy band (0.3-2.0 keV) and a hard X-ray energy band (2.0-10.0 keV) and plotted HR variations with time. 

\subsection{ \it Duty Cycle}\label{sec:DC}

Calculating the duty cycle (DC) gives us a direct estimation of the time fraction for which the object was variable. We determined the DC for Mrk 421 using the definition of \citet{1999A&AS..135..477R}, which was thereafter adopted by many authors 
\citep[e.g.,][]{2016MNRAS.455..680A}. LCs with monitoring sessions of at least 2 h were considered while doing the calculations for

\begin{equation} 
DC = 100\frac{\sum_\mathbf{i=1}^\mathbf{n} N_i(1/\Delta t_i)}{\sum_\mathbf{i=1}^\mathbf{n}(1/\Delta t_i)}  {\rm per cent} ,
\label{eq:dc} 
\end{equation} 
where $\Delta t_i = \Delta t_{i,obs}(1+z)^{-1}$ is the redshift corrected time duration for which the source was observed having the $i^{th}$ Observation ID. $N_i$ takes the value 1 whenever IDV is detected and is taken to be 0 for no IDV detection. Computation of the DC has been weighted by the actual observing run time $\Delta t_i$ on the $i^{th}$ night, as the monitoring duration is different for different observations.

\begin{figure*}
\centering
\includegraphics[height=5cm,width=17cm]{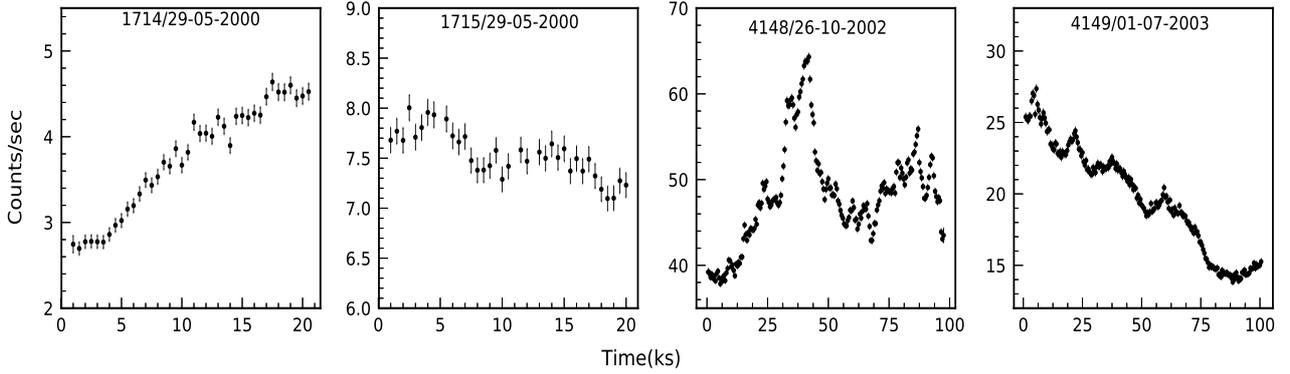}\\
\caption{{\it Chandra} sample light curves of the TeV HBL Mrk 421, labelled with Obs IDs and dates. Full figure  
appears as supplementary material on-line only.}
\end{figure*}

\begin{figure*}
\centering
\includegraphics[height=5cm,width=17cm]{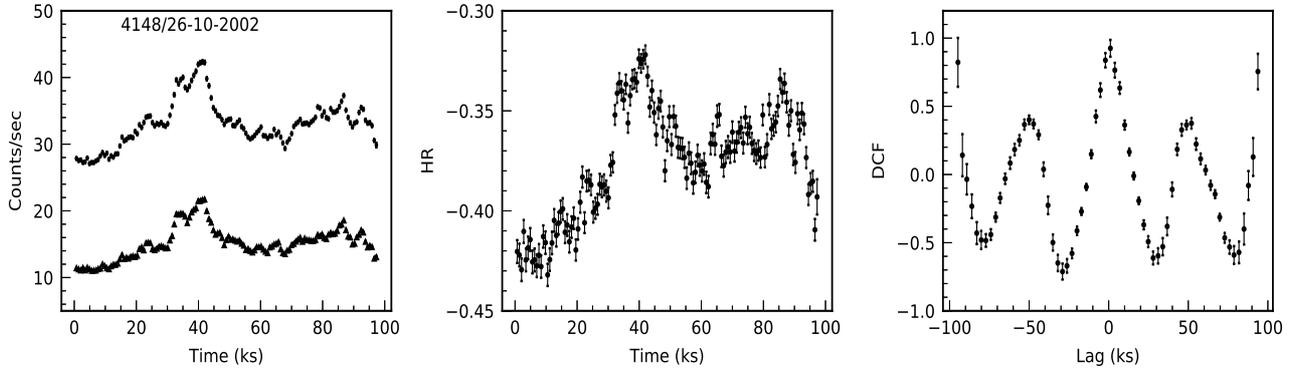}\\
\caption{Soft (0.3-2 keV, denoted by filled circles) and hard (2-10 keV, denoted by filled triangles) LCs are in the left panel, hardness ratio is in the middle panel, and the DCF  between soft and hard LCs comprises the right panel. Corresponding observation ID of Mrk 421 and the date are marked in the left panel. The observations are restricted to ACIS detectors as the HRC detector does not have an energy column and thus cannot be separated into soft and hard energy bands. Full figure appears as supplementary material on-line only.}
\end{figure*}

\begin{figure*}
\centering
\includegraphics[height=5cm,width=17cm]{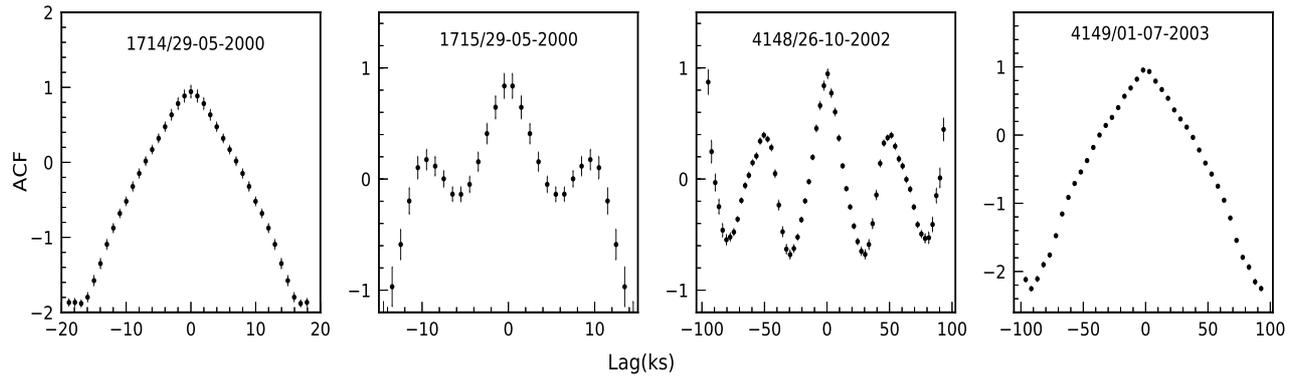}\\
\caption{Auto Correlation Function (ACF) plots for sample LCs of the blazar Mrk 421. Full figure appears as supplementary material
on-line only.}
\end{figure*}

\section{RESULTS}

We have analysed 72 {\it Chandra} observations of Mrk 421 that provide good data, as tabulated in Table 1.  The resulting X-ray LCs are shown in Fig.\ 1.  We have searched for IDV in these LCs and visual inspection of them frequently show the presence of clear variability \citep{2017MNRAS.469.3824K}. We have computed IDV variability parameters in Table 2, which include excess variance, fractional variance (F$_{var}$) and any timescales indicated by the  ACFs.  It is evident that the  F$_{var}$ amplitudes of most of the individual observations are considerable, with values exceeding 3 times the error for the total counts for 65 out of the 72 observations \citep{2015ApJ...811..143P, 2016ApJ...819..156B}.  Observation ID 11966, taken on 2010 March 14, shows the highest value of variability amplitude (21.31 $\pm$ 0.12 per cent) over the entire period, and was observed for 30 ks.  Whenever possible (for ACIS data only), we split the X-ray light curves into two energy bands: the soft X-ray energy band spans 0.3--2.0 keV while the hard band is taken as 2.0--10.0 keV. The fractional variances separately calculated for hard and soft bands are also given in Table 2; this data provides evidence for the hard band being more variable than the soft band (for 41 out of 48 observations). Statistically valid variations from Mrk 421 are clearly indicated by F$_{var}$ computations. These allowed us to compute the  DC of our source to be $\sim$ 84\%, showing a strong presence of X-ray IDV.

\par Fig.\ 2 displays the LCs of the soft and hard energy bands, both of which are shown in the left panels, with HR plots in the middle panels and DCF results between the hard and soft LCs displayed in the right panels. The hardness ratios  plotted against time show clear indications of spectral variations. 
The DCF plots quantify the strong positive correlation between the hard and soft bands, 
 at least whenever considerable variability is present. We fitted each DCF plot with a Gaussian function ($y = A {\rm exp}(-(x - x_0)^2 / w)$, with amplitude $A$, central value $x_0$, and width, $w$) to estimate any possible time lag, and found a time lag almost equal to zero in each case.
These results suggest that both soft and hard bands are emitted from the same region at the same time \citep[e.g.,][]{2017ApJ...841..123P}. 

\par The Auto-Correlation Function plots are displayed in Fig.\ 3, and they provide strong evidence of variability timescales for seven observation IDs. The timescales are taken from the locations of significant non-zero peaks in the ACF.  For seven plots such timescales range from 5.5 to 30.5 ks and are presented in Table 2. The remaining ACF plots either exhibit such a high noise level that any variability timescales cannot be ascertained or simply do not indicate the presence of any variability 
\citep{2016MNRAS.462.1508G}.

\par The course of this entire set of {\it Chandra} observations of Mrk 421 spans over 16  years.  The LTV LC of the mean of the total X-ray fluxes  is shown in Fig.\ 4.  Very substantial variations, ranging up to  21.31 $\pm$ 0.12 per cent, are seen in the individual observational count rates.  The LTV count rates vary from 0.396$\pm$0.004 to 48.226$\pm$0.023, clearly emphasizing how variable this blazar is.  No clear patterns can be discerned from a visual inspection of these LTV data.

\par In Fig.\ 5 we show plots of the HR against flux. This spectral index-flux representation was studied to uncover patterns in the HR as a function of 0.3--10.0 keV count rates for different time intervals. Spectral evolution of Mrk 421 during this $\approx$16 yrs of  observation is clearly marked by changing relative strength of particle acceleration and synchrotron cooling processes in X-ray emitting regions \citep{2015MNRAS.451.1356K}. The presence of clockwise or anti-clockwise loops in this spectral hardness-flux plots reveals information about the leading emission mechanism during that particular period. 
For these HR against count rate plots, we considered the entire set of 48 observations made with ACIS for which the HR can be measured and if we detected any loop in the plot we considered that segment as one epoch and started a search for subsequent loops. This way, we found eight distinct epochs having either clockwise or anti-clockwise loops that include a substantial majority of the observations (31 of 48).
X-ray emissions  can  be understood as arising from high energy relativistic particles that are accelerated as a result of shocks propagating in the relativistic jets. When these accelerated particles come in contact with the inhomogeneous magnetic fields significant synchrotron emission extends into the X-ray band, which will dominate the cooling process. Fig.\ 5 contains an anti-clockwise loop for Epoch 1 which can be understood as a hard-lag \citep{2002ApJ...572..762Z}. This leads us to conclude that particles were being accelerated to very high speeds during that time span \citep{2015MNRAS.451.1356K}. Dominance of particle acceleration  mechanisms were again indicated in Epoch 3. Epochs 4 and 7  also have anti-clockwise loops, providing  evidence for particles being accelerated in the internal shocks as they outflow along the jets. Clockwise loops, or soft-lags, were displayed by Epochs 2 and 5 and then again by Epochs 6 and 8. This implies that the synchrotron cooling mechanism dominated the X-ray emission in those particular periods. 

\section{DISCUSSION}

\begin{figure}
\centering
\includegraphics[width=90mm]{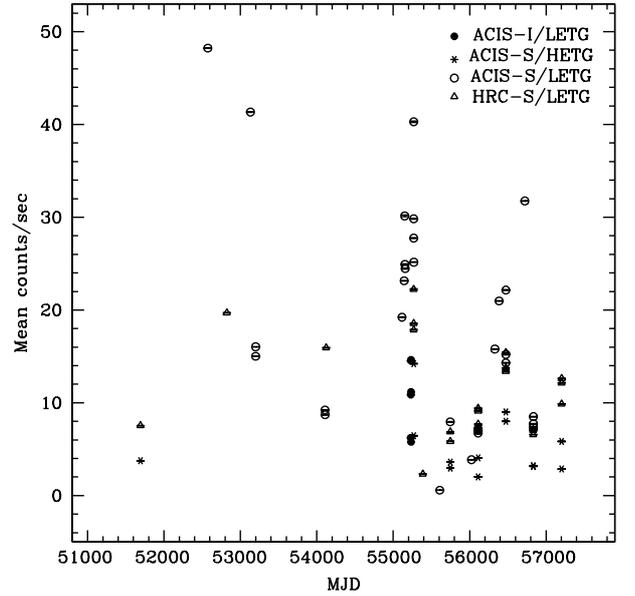} 
\vspace*{-0.3in}
\caption{Long-term X-ray variability in Mrk 421.}
\end{figure}

An important route to understanding the emission mechanisms in blazars and other AGN involves careful measurement of strong flux variations on diverse timescales. Studies of rapid variability of blazars have also helped us to determine the key properties of the emitting region such as its size, location and structure \citep[e.g.,][]{2003A&A...400..487C, 1976ARA&A..14..173S, 2016MNRAS.462.1508G}. Whenever extreme variability is combined with relatively weak spectral features, as in BL Lac objects,  it has long been accepted  that a relativistic jet close to the line of sight (LOS) is emitting the continuum; this  results in the dramatic increase in the brightness of the observed radiation and a reduction of the observed variability timescales due to Doppler boosting \citep{1978PhyS...17..265B, 1997ARA&A..35..445U}. The high X-ray variations displayed by many blazars could arise either directly from the synchrotron emission  or through Compton scattering of the lower-energy synchrotron photons (SSC) process, again supporting the idea of relativistic bulk motion in blazars \citep{1966ApJ...144..534H, 1983SSRv...35..367B}.  In general, intrinsic AGN emission and variability can occur through  two fundamental theoretical branches, either the purely relativistic-jet-based models 
\citep{1985ApJ...298..114M, 1992A&A...259..109G, 2014ApJ...780...87M, 2015JApA...36..255C} or the accretion-disc based models 
\citep{1993ApJ...406..420M, 1993ApJ...411..602C}.  IDV and STV in  radio-quiet quasars, and perhaps in certain blazars, if they are in very low states, can possibly be explained through instabilities present in accretion discs \citep{1993ApJ...406..420M}. Yet any such accretion-disc based model does not provide a satisfactory explanation for  the very strong and rapid variability on most timescales that can be much more easily provided by the doppler boosted radiations from relativistic jets \citep{1993ApJ...411..602C}.

\begin{figure*}
\centering
\includegraphics[width=170mm]{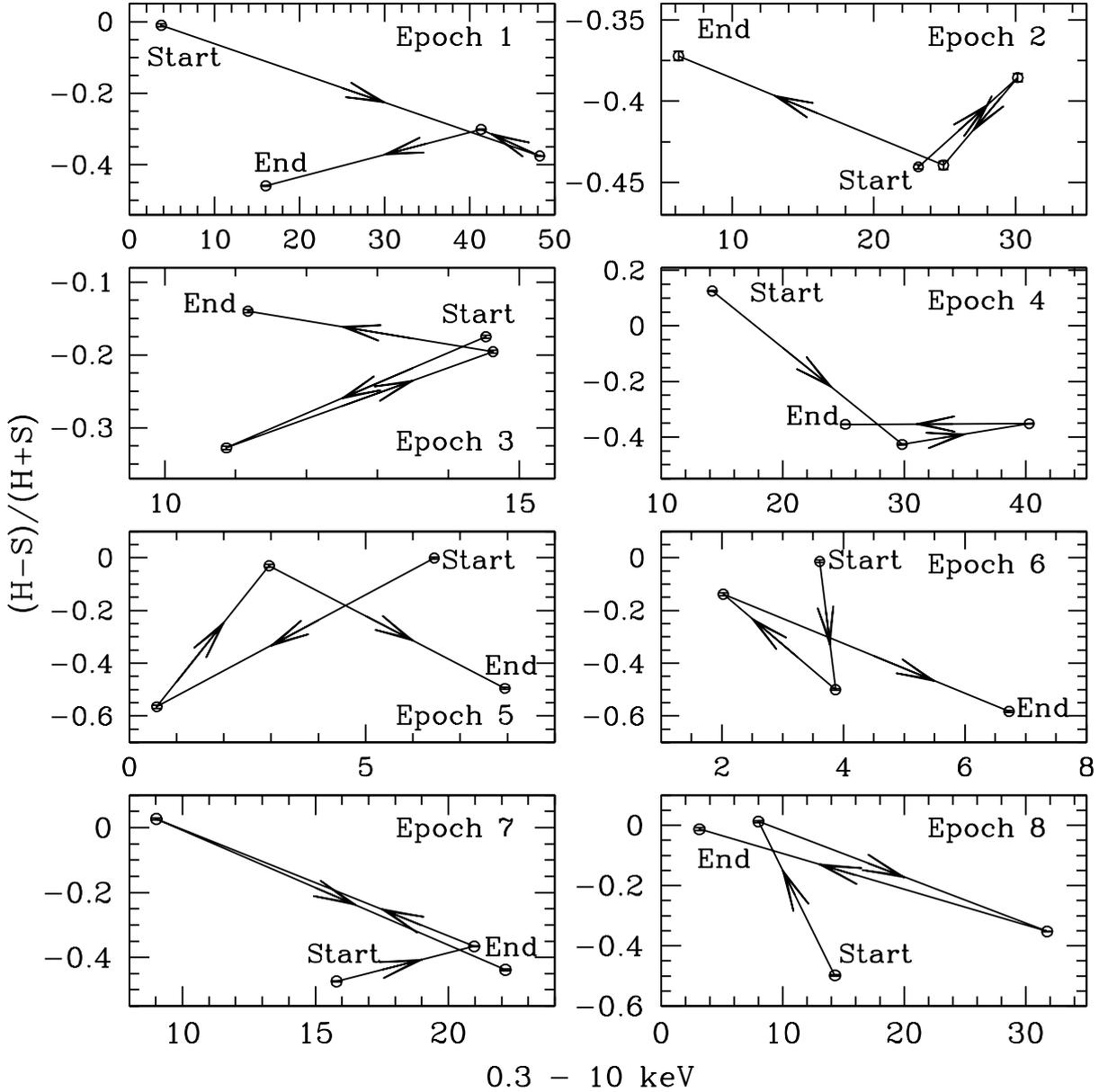} 
\caption{Spectral variations of Mrk 421 in various epochs with start and end points marking the loop directions. Each epoch corresponds to the time interval during which the data were acquired for each loop, considered from Epoch 1 to Epoch 8: Epoch 1: 29-05-2000 to 13-07-2004; Epoch 2: 08-11-2009 to 02-02-2010; Epoch 3: 04-02-2010 to 06-02-2010; Epoch 4: 13-03-2010 to 14-03-2010; Epoch 5: 14-03-2010 to 04-07-2011; Epoch 6: 05-07-2011 to 03-07-2012; Epoch 7: 07-02-2013 to 30-06-2013; and Epoch 8: 01-07-2013 to 25-06-2014.}
\end{figure*}

\par An adiabatic shock-in-jet model can nicely explain blazar variability on LTV timescales \citep{1985ApJ...298..114M, 1995ARA&A..33..163W}. The model describes relativistic shocks as arising from disturbances created in the inner portion of the jet that can quickly steepen into shocks. As they propagate through the jet, these shocks create major flux.  Additional fluctuations are observed when these shocks interact with helical jet structures, by changing the effective viewing angle and hence Doppler factor  
\citep{1992A&A...259..109G, 1992A&A...255...59C, 2017ApJ...841..123P}. Relativistic jets also suffer some turbulence behind some of the  shock regions, which can be held responsible for smaller STV and IDV \citep{2014ApJ...780...87M, 2015JApA...36..255C, 2016ApJ...820...12P}. 

Some blazars, such as Mrk 421, exhibit flaring TeV emissions on the timescales of a few minutes, which can be short when compared to the light crossing time of the SMBH of those blazars. Also, for TeV seed photons to escape from their source region, which is quite compact, it appears that the Lorentz factor of the emitting region needs to be $\gtrsim$ 50. This is required in order to avoid absorption via pair production through interaction with soft radiation fields, but such extreme Lorentz factors are hard to achieve with a jet with a uniform bulk flow.  The flaring states of Mrk 421 might be more naturally explained via a "jets-in-a-jet" model proposed by \cite{2009MNRAS.395L..29G}.  This scenario can explain the origin of TeV emission along with fast X-ray variability though production of rapid flares arising from to the ultrarelativistic outflow of material from magnetic reconnection sites.  The presence of multiple reconnection regions and the phenomena related to the rupturing of a large reconnection site make this model more flexible, so that the production of more slowly varying flares is also possible. Electrons accelerated within the jets would be responsible for rapid hard X-ray variability and the successive flares detected in Mrk 421. This phenomena can also yield  the observed spectral hardening at high energies.

\par Certain parameters can be  estimated, rather independently of  model details, by assuming that synchrotron emission is responsible for the X-ray emission from HBLs, such as Mrk 421. In the observer's frame, the acceleration timescale of the  diffuse shock acceleration  mechanism \citep{1987PhR...154....1B} assumed to be the way in which  electrons are accelerated, is given as \citep{2002ApJ...572..762Z}
\begin{equation}
t_{acc}(\gamma) \simeq 3.79\times10^{-7} \frac{(1+z)}{\delta} \xi B^{-1}\gamma ~{\rm s},
\end{equation}
where, $\xi$ is defined as the acceleration parameter, $B$ is the magnetic field and $\gamma$ is the Lorentz factor of the electrons. Synchrotron emission is also believed to be the origin of the X-ray emission, when TeV blazars are considered. An individual electron with energy ${\it E = \gamma m_{e}c^{2}}$ has a synchrotron cooling timescale of,
\begin{equation}
t_{cool}(\gamma) \simeq 7.74\times10^{8}\frac{(1+z)}{\delta} B^{-2}\gamma^{-1} ~{\rm s},
\end{equation}
For the {\it Chandra} energy range, the critical synchrotron emission frequency $\nu \simeq$ 4.2$\times$$10^{6}$$\frac{\delta}{1+z}$B$\gamma^{2}$ = $10^{18}\nu_{18}$ Hz.
\par Although the minimum variability timescale found in this work is 5.5 ks, the variability associated with that observation is only $\sim$3\% so we took 9.51 ks as the shortest clear variability timescale, as $F_{var}$ exceeds 7\% for it. The cooling timescale should be greater than or equal to the minimum variability timescale or $\sim$ 9.51 ks for Mrk 421 in this work. This implies,
 
\begin{equation} 
B \geq 0.30(1+z)^{1/3}\delta^{-1/3}\nu^{-1/3}_{18} ~{\rm G}.
\end{equation}
Although a range of values for $\delta$ for Mrk 421 between 20 and 50 \citep[e.g.,][and references therein]{1998ApJ...509..608T,2011ApJ...736..131A,2016MNRAS.463.4481Z} appear in the literature, 
we choose  $\delta$=25 \citep[e.g.,][]{2016ApJ...819..156B}, and get,
\begin{equation}
B \geq 0.10~ \nu^{-1/3}_{18} {\rm G}.
\end{equation}
Combining these values of $B$ and $\delta$, we estimate the electron Lorentz factor as,
\begin{equation}
\gamma \geq 3.06\times10^{5} ~\nu^{2/3}_{18}.
\end{equation}
The characteristic radius of emitting region can also be evaluated using the bound
\begin{equation}
R \leq \frac{c ~t_{var} ~\delta}{1+z} \leq 6.92\times10^{15} {\rm cm}.
\end{equation}

Relativistic electrons directly responsible for hard, variable X-ray emission, as are investigated by {\it Chandra}, must be repeatedly accelerated because of the short cooling timescales of these very high-energy electrons \citep{2017ApJ...841..123P}. These electrons must be injected via one or more of the various possible acceleration mechanisms mentioned above.  We note that diffusive-shock acceleration could  be responsible both for variations in the flux and spectral hardening at high energies.

For this HBL we probed the X-ray spectral variability by analyzing the HR, as it serves as an easy and efficient way to understand changes in the spectra; however, the  physical parameters that are responsible for spectral variability are not directly evaluated by this method.  We observed that for Mrk 421  the HR increases as the flux (count rate) increases, or, as the flux is increasing, the spectrum tends to get flatter. Thus it seems that this source is following the general trend of ``hardens-when-brighter" of the HSP type blazar as discussed previously \citep{1998ApJ...492L..17P, 2002ApJ...572..762Z, 2003A&A...402..929B, 2004A&A...424..841R, 2017ApJ...841..123P}. The spectral hardness--flux analysis we conducted provides a model independent approach to understanding the spectral variations of the source. The presence of anti-clockwise loops in the plots indicates that the soft band leads the hard band (or there is a hard-lag between the two emissions), i.e., with the increase in the total flux of the source, the hard flux usually increases more than the soft flux. This indicates that the particle acceleration mechanism is predominantly responsible for the observed X-ray emission during that particular period of time \citep{2016NewA...44...21B}. When any clockwise loop (or soft-lag) occurs in the  HR against flux plot, the soft flux increases more than the hard flux with increasing total flux. In these epochs, the hard band leads the soft band, as is the case when the synchrotron emission mechanism temporarily dominates the acceleration mechanism.

\section{CONCLUSIONS}

We studied  72 {\it Chandra} light curves of the TeV blazar Mrk 421, and searched for variability timescales of IDV. The rapid X-ray variability studied here most likely originates within  compact regions of the relativistic jet. Our conclusions are summarised as follows:
\begin{enumerate}
  \item The fractional variability amplitude provides a clear indications of variability on many occasions, with highest variability amplitude being over 21 per cent. The variability in hard energy X-ray bands  presumably originates from a compact region within a relativistic jet.
  \item The duty cycle for these variations is at least $\approx$ 84 per cent which indicates that the source was exceptionally variable in the observed 16 year span.
 \item We found evidence for timescales ranging from 5.5 to 30.5 ksec in 7 LCs of Mrk 421 using the ACF technique. Other observations have noisier ACF plots in which variability timescales are not clearly present. 
Using the shortest strong variability timescale of 9.51 ks,  we can estimate key parameters in a fashion this is essentially independent of the theoretical model.  We find a magnetic field  $B \textgreater 0.10 ~\nu^{-1/3}_{18}$  G, electron Lorentz factor $\gamma \geq 3.06 \times10^{5} \nu^{2/3}_{18}$ and radius of the emitting region $R  \leq 6.92 \times 10^{15}$ cm.
 \item The DCF technique was applied to the hard (2--10 keV) and soft (0.3--2 keV) X-ray bands and displayed positive correlations with no time lag. This implies that the emission in both nearby bands arose from the same production region at the same time; i.e., there is no evidence that the softer X-rays arise from synchrotron emission while the harder come from SSC.
 \item A hardness ratio analysis was also employed to study spectral variations.  HRs   showed flatter spectra at high fluxes (Fig.\ 2). This indicated that the HR normally increased with increasing flux and got ``harder-when-brighter". Fig.\ 5 displays more information about the spectral evolution of the source. During these extended observations, a few epochs had hard-lags, pointing to the particle acceleration mechanism being responsible for the X-ray emissions, whereas a few epochs had soft-lags,  indicating that X-rays are predominantly emitted by the synchrotron cooling mechanism during those periods.
 \end{enumerate}

\section*{Acknowledgements}
We thank the referee for constructive comments and suggestions.
This research has made use of data obtained from the Chandra Data Archive and the Chandra Source Catalog, and software provided by the Chandra X-ray Center (CXC) in the application packages CIAO, ChIPS, and Sherpa. 

The work of VA, ACG and KKY is supported by BRNS-DAE (Board of Research in Nuclear Sciences -- Department of Atomic Energy), Government of India funded project No. 37(3)/14/17/2014-BRNS. ZZ is thankful for support from the Chinese Academy of Sciences Hundred-Talented program (Y787081009).  
PJW is grateful for hospitality at the Kavli Institute for Particle Astrophysics and Cosmology, Stanford University, and Shanghai Astronomical Observatory during a sabbatical.

\bibliographystyle{mnras}
\bibliography{master}

\end{document}